\documentclass[useAMS,usenatbib,a4paper]{mn2e}
\usepackage{amsmath}
\usepackage{graphicx}
\usepackage{aas_macros}
\usepackage[T1]{fontenc}
\usepackage{aecompl}
\newcommand{\bb}{\bmath{\beta}}

\newcommand{\bo}{\bmath{\omega}}
\newcommand{\bp}{\bmath{\phi}}
\newcommand{\bM}{\mathbf{M}}
\newcommand{\bt}{\bmath{\theta}}
\newcommand{\vv}{\bmath{v}}
\newcommand{\bbt}{\bmath{\bar\theta}}
\newcommand{\dbt}{\Delta\bmath{\theta}}
\newcommand{\dt}{\Delta\theta}
\newcommand{\tE}{\theta_{\textsc{e}}}

\newcommand{\Bxi}{\bmath{\xi}}

\DeclareMathOperator{\var}{var}
\title[Microlensing of QSOs]{A method for the microlensed flux variance of QSOs}
\author[J. J. Goodman and A.-L. Sun]{Jeremy Goodman$^{1}$ and Ai-Lei Sun $^{1}$ \\
$^1$Department of Astrophysical Sciences, Princeton University, Princeton, NJ 08544, USA}

\begin{document}
\date{Drafted 2013 August 27.}
\pagerange{\pageref{firstpage}--\pageref{lastpage}} \pubyear{2013}
\maketitle
\label{firstpage}

\begin{abstract}
  A fast and practical method is described for calculating the microlensed flux variance of an
  arbitrary source by uncorrelated stars.  The required inputs are the mean convergence and shear
  due to the smoothed potential of the lensing galaxy, the stellar mass function, and the absolute
  square of the Fourier transform of the surface brightness in the source plane.  The mathematical
  approach follows previous authors but has been generalized, streamlined, and implemented in
  publicly available code.  Examples of its application are given for Dexter and Agol's
  inhomogeneous-disk models as well as the usual gaussian sources.  Since the quantity calculated is
  a second moment of the magnification, it is only logarithmically sensitive to the sizes of very
  compact sources.  However, for the inferred sizes of actual QSOs, it has some discriminatory power
  and may lend itself to simple statistical tests.  At the very least, it should be useful for
  testing the convergence of microlensing simulations.
\end{abstract}


\section{Introduction}
\label{sec:intro}
Gravitational microlensing by stars along the line of sight to a QSO is sensitive to the size and
structure of the optically luminous regions of the accretion disk, which are otherwise unresolvable
at present: sources of angular size much smaller than the Einstein ring of a lensing star can be
strongly amplified, whereas more extended sources cannot be \citep[e.g.,][]{Young1981}.  Recent
quasar surveys have turned up a number of quasars suitably aligned with intervening galaxies, and
the analysis of their light curves has yielded two principal results.  Firstly, in most cases the
source size scales with wavelength approximately as expected for a steadily accreting, optically
thick disk, namely $\theta\propto\lambda^{4/3}$
(\citealt{Anguita+etal2008,Eigenbrod+Courbin+Meylan+Agol2008,Poindexter+Morgan+Kochanek2008,
  Bate+etal2008,Mosquera+etal2011, Munoz+etal2011,Mosquera+etal2011}; but
\citealt{Floyd+Bate+Webster2009} and \citealt{Blackburne+etal2011} find otherwise).  Secondly
however, the absolute source size is too large by at least half an order of magnitude
\citep{Pooley+Blackburne+Rappaport+Schechter2007,Blackburne+etal2011,
  Jimenez-Vicente+Mediavilla+Munoz+Kochanek2012}.  Not only is the source larger than thin-disk
theory predicts for likely ranges of black-hole mass and accretion rate, but also, as
\cite{Morgan+etal2010} have emphasized, it is too large for any source that radiates locally as a
black body unless the radial temperature profile is substantially shallower than $\theta^{-3/4}$.

If the latter conclusion is correct, then something is seriously wrong with steady-state thin-disk
theory as applied to QSOs.  The discrepancy is not small; stated in terms of the areas rather than
the linear sizes of the sources, it is more than an order of magnitude.  Several physical
possibilities would need to be explored, ranging from highly inhomogeneous disks
\citep{Dexter+Agol2011}, perhaps caused by thermal or viscous instabilities
\citep{Lightman+Eardley1974,Sunyaev+Shakura1975}; to disk warps, perhaps driven by a
radiation-pressure instability \citep{Pringle1996}; to optically thick scattering-dominated winds;
or even gap opening by embedded satellites \citep[e.g.,][]{Armitage+Natarajan2002, Goodman+Tan2004}.
However, the error bars in the microlensing size estimates are still large; studies differ as to the
magnitude of the size discrepancy, and sometimes even the sign \citep{Rauch+Blandford1991}.  The
statistical methods used sometimes opaque, especially when entire photometric time series are fitted
to simulations.  For these among other reasons, one worries that the overlarge inferred sizes may be
due to inadequate statistics or systematic errors.  Ultimately this will be decided only by more
data independently analyzed.

In the meantime, having no data of our own to offer, we revive and extend a
semianalytic approach to the prediction of microlensing statistics pioneered by
\cite{Deguchi+Watson1987} and improved by \cite{Seitz+Schneider1994},
\cite{Seitz+Wambsganss+Schneider1994}, and \cite{Neindorf2003}.  The semianalytic method adds no
noise to the comparison between models and data and avoids possible biases due to choices of numerical
parameters, such as the numbers of rays shot or the size of the regions simulated in the lens and
source plane.  It is easily extended to the autocorrelation of the magnification as a function of
time lag provided that the velocity dispersion of the lensing stars is small compared to the motion
of the lensing galaxy across the line of sight.  The method is limited, however, to second moments
of the light curve.  Simulations are more flexible and can address higher moments of the
magnification and the structure of caustic crossings, effects that are more sensitive to the sizes
of very compact sources.  These are good scientific reasons to prefer simulations.  Certain
practical impediments to the use of the semianalytic method, however, are removable.  The method
is perceived to be cumbersome and hard to use, involving as it does
multiple integrals and expansions in special functions.  In the efficient but restricted
version implemented by \cite{Neindorf2003}, it is limited to gaussian sources.
In this paper, we develop an efficient and practical version of the semi-analytic formalism that can
be applied to arbitrary source structures, not just gaussian ones.  In hopes
that it will be more widely used, we have implemented the method in {\tt Python} code downloadable
from a {\tt git} repository {\tt https://bitbucket.org/jjgoodman/mulensvar}.  

The effort required by this method factors in a way well suited to exploring a large range of source
structures and macrolensing parameters.  The main part of the work is to calculate a kernel that
depends upon dimensionless statistical properties of the lens such as optical depth, shear, and mean
stellar mass or mass distribution.  This kernel can be computed efficiently for almost any
reasonable choice of these parameters, using a pre-computed special function (\S\ref{subsec:inner}).
The kernel so obtained is
independent of the source structure.  Second moments of magnification are obtained by convolving the
kernel with the spatial autocorrelation function of the source.  Hence if one is reasonably
confident of the average properties of the lenses, then a single computed kernel may quickly be
applied to many possible sources.  Even if simulations are necessary, one can at least use this
method as partial check on convergence with respect to the purely numerical parameters of the
simulations, such as the number of rays, stars, and realizations.  Although agreement with the
semianalytic flux variance does not guarantee that the simulations are fully converged for all
purposes, one should not trust simulations that do not show such agreement.

The plan of this paper is as follows.  \S\ref{sec:definitions} introduces the lensing kernel and
associated notation.  \S\ref{sec:evaluation} describes our algorithms for evaluating the lensing
kernel numerically and (in certain limits) analytically.  Most of this quite technical section can
safely be skipped if one wishes to use our codes as they are, but one should read it carefully
before attempting modifications.  The asymptotic analytic results \eqref{eq:Jsmallom} \&
\eqref{eq:Jlargeom} and their physical interpretation may be of more general interest, however.
Interpolation between these formulae captures much of the behavior of the microlensing kernel, as is
demonstrated by Fig.~\ref{fig:fig3}.  \S\ref{sec:examples} presents numerical examples and tests.
These include applications to Huchra's Lens (Q2237+305) and to \cite{Dexter+Agol2011}'s fragmented
disks.  The predictions of the method are compared with simulations made using a simple ray-shooting
code.  The agreement is satisfactory, but the convergence of the simulations to the predictions is
somewhat slow with regard to domain size and number of stars: regions $\ga 10^2\tE$ on a side are
needed for $1\%$ accuracy.

\section{Definitions}
\label{sec:definitions}

Macrolensing by the large-scale mean potential of the intervening galaxy splits the source into
macroimages separated typically by arcseconds.  Microlensing by stars divides each macroimage into a cluster of
subimages with angular separations comparable to the Einstein ring of a single star, $\tE = \sqrt{4GM_*D_{LS}/c^2 D_S
  D_L}$, which is $\approx 2\,(M_*/M_\odot)^{1/2} \,{\rm \umu as}$ ($\approx 10^{-11}\,{\rm rad}$) for typical lens and source
redshifts $(z_L,z_S)=(0.5,2.0)$.  At the angular-diameter distance of the lens, this projects to $D_L\tE\sim
10^{-2}\,{\rm pc}$.  In theoretical studies of microlensing, it is often necessary to consider a region several tens of
$\tE$ across to allow for the interactions between caustics and for the motion of the line of sight across the lens
plane (at, say, $0.03\tE\,{\rm yr^{-1}}$).  Even so, the region of interest is scarcely larger than a parsec.  On such
scales the contribution of the large-scale smoothed galactic potential to the lens equation is well approximated by a
locally constant linear transformation,
\begin{equation}
  \label{eq:Mmatrix}
  \bM =
  \begin{pmatrix}
    1-\kappa-\gamma\cos 2\alpha & -\gamma\sin 2\alpha \\
    -\gamma\sin 2\alpha & 1 -\kappa +\gamma\cos 2\alpha
  \end{pmatrix}.
\end{equation}
The notation of \cite{Miralda-Escude1991} is used for the dimensionless convergence $\kappa$, total
shear $\gamma$, and orientation $\alpha$ of the principal axes of the shear. The macrolensing
magnification is $(\det\bM)^{-1}=[(1-\kappa)^2 -\gamma^2]^{-1}$.  The local mean surface density in
stars contributes a portion $\kappa_*$ to the convergence.  The remainder,
$\kappa'=\kappa-\kappa_*$, is due to dark matter and gas and is assumed to be smoothly distributed.
Typically $\kappa_*/\kappa\sim 0.05-0.1$ \citep{Mediavilla+etal2009}, but stars may dominate the
convergence when the macroimage lies in the central parts of the lensing galaxy, as in Q2237+305.

Within a single macroimage, the lens equation relating a line of sight from Earth in direction $\bt$ to its
unlensed counterpart $\bt_S$ on the source plane becomes
\begin{equation}
  \label{eq:lenseqn1}
  \bmath{\theta}_S = \bM\cdot\bt+ \left(\tE^2
\sum_k \frac{\Bxi_k-\bt}{|\Bxi_k-\bt|^2}  + \kappa_*\bt\right) \equiv \bM\cdot\bt + \bp\,.
\end{equation}
Here $\Bxi_k$ is the angular position of the $k^{\rm th}$ microlensing star, with the origin of
coordinates for $\bt$ and $\Bxi_k$ taken at the center of the macroimage.  The term $\kappa_*\bt$
has been grouped with the sum over stars to prevent double-counting the mean stellar convergence,
since $-\kappa=-\kappa_*-\kappa'$ in $\bM$.  For the moment, all stars have the same mass and
therefore the same $\tE^2$, but a mass function will be introduced later (\S\ref{subsec:mspec}).  In
the approximation that the sum above ranges over stars distributed throughout the lens plane with
constant mean number per unit area, the net stellar deflection $\bp$ is a random variable with zero
mean and stationary statistics.  In other words, the joint probability density for this variable to
take on values $\{\bp_1,\ldots\bp_n\}$ at positions $\{\bt_1,\ldots\bt_n\}$ depends upon the
differences $\bt_i-\bt_j$ but not on the centroid $\bmath{\bar\theta} = (\bt_1+\ldots+\bt_n)/n$.
Binary and higher-multiplicity correlations among the stars would not
be incompatible with stationary statistics but would make results along the lines of this
paper almost impossible.   Such correlations could be represented in simulations quite easily but
rarely are.

An explicit expression is possible for
Fourier transform $P(\bo_1,\ldots,\bo_n)$ of the $n$-point probability density
$P(\bp_1,\ldots\bp_n)$.  We call this the ``characteristic function:''
\begin{multline}
  \label{eq:charfct}
  \left\langle\exp\left[i\bo_1\cdot\bp(\bt_1)+\ldots+i\bo_n\cdot\bp(\bt_n)\right]\right\rangle= \\
\exp\left\{\nu\int\left[
\exp\left(i\sum_{j=1}^n\frac{\tE^2\bo_j\cdot(\Bxi-\bt_j)}{|\Bxi-\bt_j|^2}\right) \right. \right.\\
\left.\left.-1-i\sum_{j=1}^n\frac{\tE^2\bo_j\cdot(\Bxi-\bt_j)}{|\Bxi-\bt_j|^2}\right]\,d^2\Bxi \right\}\,.
\end{multline}
Here $\langle\ldots\rangle$ denotes expectation value, and $\nu\equiv\kappa_*/\upi\tE^2$ is the mean
number of stars per unit area.  \cite{Chandrasekhar1943} derived this for $n=1$, and
\cite{Deguchi+Watson1987} stated the result for $n=2$ without derivation. \cite{Seitz+Schneider1994}
devote an entire paper to the discussion of the 2-point function and its inverse Fourier transform.
For completeness, we sketch a derivation here for general $n$: Consider the counterpart to the left
side of eq.~\eqref{eq:charfct} when $\bp_j$ replaced by $\bp'_j\equiv\bp_j-\kappa_*\theta_j$.  The
primed variables do not have stationary statistics because their means are $-\kappa_*\bt_j$.
However, the contribution of a small area $A$ of the lens plane with centroid $\Bxi_{ A}$ to their
characteristic function is, with $f_{ A}\equiv \sum_j \tE^2\bo_j\cdot(\Bxi_{ A}-\bt_j)/|\Bxi_{
  A}-\bt_j|^2$,
\begin{equation*}
\sum_{N_{ A}=0}^\infty \frac{(\nu A)^{N_{ A}}}{N_{ A} !} e^{-\nu A} e^{i N_{ A} f_{ A}}
=\exp\left[\left(e^{if_{ A}}-1\right)\nu A\right]\,,
\end{equation*}
if the number of stars within area $A$ ($N_A$) is Poissonian with mean $\nu A$.  Multiplying the
independent contributions from all such areas in the lens plane gives eq.~\eqref{eq:charfct} except
for the final term within the square brackets, which can be explained as follows: If the stars
were restricted to a circular region $|\Bxi| < R$, then the average deflection due to these stars at
$|\bt_j|<R$ would be
\begin{equation*}
\nu\tE^2\int\frac{\Bxi-\bt_j}{|\Bxi-\bt_j|^2}\,d^2\Bxi = -\kappa_*\bt_j\,.
\end{equation*}
Thus the term in question removes the mean deflection from the characteristic
function, as appropriate for the $\bp_j$ rather than the $\bp'_j$.

\subsection{Moments of the flux}
\label{subsec:moments}

For a transparent lens, the observed surface brightness $I$ in direction $\bt$ is equal to the
unlensed surface brightness $I_0$ at the position $\bt_S=\bM\cdot\bt+\bp(\bt)$ where the lensed ray
intercepts the source plane.  Therefore, the flux of a macroimage summed over all of its microimages
is
\begin{equation*}
  F = \int I(\bt)\,d^2\bt  = \int I_0(\bM\cdot\bt+\bp)\,d^2\bt\,,
\end{equation*}
The source and lensing variables can be separated by introducing the Fourier transform $\hat I_0$ of the
unlensed source,
\begin{equation}
  \label{eq:FTI}
I_0(\bt) = \int \frac{d^2\bo}{(2\pi)^2} \hat I_0(\bo) e^{i\bo\cdot\bt}\,,
\end{equation}
so that
\begin{equation}\label{eq:flux}
    F = \int \frac{d^2\bo}{(2\upi)^2}\hat I_0(\bo) \int d^2\bt\,   e^{i\bo\cdot(\bM\cdot\bt+\bp)}\,.
\end{equation}
The $n^{\rm th}$ moment of the flux is therefore
\begin{align*}
  \langle F^n\rangle = & \left[\prod\limits_{j=1}^n\int \frac{d^2\bo_j}{(2\upi)^2}\hat I_0(\bo)\int d^2\bt_j
    e^{i\bo_j\cdot\bM\cdot\bt_j}\right] \\
&\times\left\langle\exp\left[i\bo_1\cdot\bp(\bt_1)+\ldots+i\bo_n\cdot\bp(\bt_n)\right]\right\rangle\,.
\end{align*}
The expectation value $\langle\ldots\rangle$ is  independent of the centroid
$\bbt\equiv (\bt_1+\ldots+\bt_n)/n$.  This can be used to reduce the number of vectorial integrations by two.
Setting $\bt_j=\bt'_j+\bbt$ and integrating over $\bbt$ produces (dropping the primes hereafter)
\begin{multline}
  \label{eq:Fn0}
    \langle F^n\rangle = (\det\bM)^{-1}\left[\prod\limits_{j=1}^n\int \frac{d^2\bo_j}{(2\upi)^2}\hat I_0(\bo)\int d^2\bt_j
e^{i\bo_j\cdot\bM\cdot\bt_j}\right]\\
\times\, (2\upi n)^2\delta^2(\bo_1+\ldots+\bo_n)\delta^2(\bt_1+\ldots+\bt_n)\\
\times\, \left\langle\exp\left[i\bo_1\cdot\bp(\bt_1)+\ldots+i\bo_n\cdot\bp(\bt_n)\right]\right\rangle\,.
\end{multline}
For $n=1$, the delta functions
absorb both integrations, so that $\langle F\rangle \to (\det\bM)^{-1}\hat I_0(\bmath{0})$, which is
the correct expression for the mean macrolensed flux.

As in previous works, our numerical methods are limited to the lowest nontrivial moment,
$n=2$.  We set $\bo_1=-\bo_2\equiv\bo$ and $\bt_2=-\bt_1 = \dbt/2$.  Then
\begin{equation}
  \label{eq:Fsq}
  \langle F^2\rangle = (\det\bM)^{-2}\int\frac{d^2\bo}{(2\upi)^2}\,|\hat I_0(\bo)|^2 \hat J(\bo),
\end{equation}
\begin{equation}
  \label{eq:Jhat}
  \hat J(\bo) = \det\bM\int d^2\dbt\,
e^{i\bo\cdot\bM\cdot\dbt}\left\langle e^{i\bo\cdot[\bp(0)-\bp(\dbt)]}\right\rangle\,.
\end{equation}
Once the kernel $\hat J(\bo)$ has been calculated, equation \eqref{eq:Fsq} can be used to find
the second moment of the flux for a
general source structure with spatial power spectrum $|\hat I_0(\bo)|^2$.  Furthermore, if one
adopts the ``frozen-screen'' approximation in which the motions of the stars within the lensing
galaxy are neglected compared to the transverse motion $\bmath{V}_\perp$ of the lensing galaxy across
the line of sight, then one can use the same kernel to calculate the correlation between the
microlensed flux at finite time lag:
\begin{align}
  \label{eq:Fcorr}
  \langle F(t)F(t+\tau)\rangle \approx  & (\det\bM)^{-2} \\
  \times& \int \frac{d^2\bo}{(2\upi)^2} |\hat I_0(\bo)|^2\ 
\hat  J(\bo)\exp\left(\frac{i\bo\cdot\bmath{V}_\perp\tau}{D_L}\right)\,. \nonumber
\end{align}
As discussed by \cite{Kundic+Wambsganss1993}, however, the frozen-screen approximation is not very
accurate, especially at high optical depth \citep{Wyithe+Webster+Turner2000}.

As discussed in \S\ref{subsec:outer}, $\hat J(\bo)$ contains a term proportional to $\delta^2(\bo)$
that accounts for the square of the mean flux in eqs.~\eqref{eq:Fsq} \& \eqref{eq:Fcorr}.  If this
term is subtracted from $\hat J(\bo)$, the right sides of these equations yield the variance and
covariance of the flux.  In other words, subtracting the delta function from $\hat J(\bo)$ has the
same effect as removing the mean from $F$ before its second moments are calculated. The numerical
method that we use to evaluate $\hat J(\bo)$ makes this subtraction automatic.  At the risk of
confusion, {\it we hereafter interpret $\hat J(\bo)$ as the modified kernel so that $\langle
  F^2\rangle\to\var(F)$ in eq.~\eqref{eq:Fsq}}.

\subsection{Transformation and scaling of the characteristic function}
\label{subsec:scaling}

In the approach outlined above, the first step toward $\langle F^n\rangle$ is to calculate the
characteristic function \eqref{eq:charfct}.  This involves integration over the vector $\Bxi$.
The result will evidently be a function of the $2n$ vector-valued
parameters $(\bo_1,\ldots,\bo_n)$ and $(\bt_1,\ldots,\bt_n)$.  Because of the two delta functions in
eq.~\eqref{eq:Fn0}, only $2n-2$ of these parameters are independent, equivalent to $4n-4$
independent scalar parameters.  Because the characteristic function is independent of the
macrolensing shear, it is statistically isotropic and hence unaffected by an overall rotation of the
lens plane.  This reduces the number of essential scalar parameters by one.  The following scaling
property reduces the number by one more.  Let $\bo_j\to\sigma\bo_j$, $\bt_j\to\sigma\bt_j$, with the
same factor $\sigma$ for all $j$.  Rescaling the dummy integration variable $\Bxi\to\sigma\Bxi$ and
the stellar number density $\nu\to\sigma^{-2}\nu$ [equivalently $\tE^2\to\sigma^{-2}\tE^2$ or
$\kappa_*\to\sigma^{-2}\kappa_*$] restores the integration to its form for $\sigma=1$.

As a result, the number of essential scalar parameters needed to specify the $n$-point
characteristic function is $4n-6$.  (This assumes $n>1$, else the characteristic function collapses
to a constant.)  For $n=2$, this is manageable.  The logarithm of the two-point characteristic
function can be tabulated for a grid of values of two parameters, each entry in the table requiring
a two-dimensional quadrature.  Fortunately the calculation need be done only once, since it is
independent of the macrolensing matrix ($\bM$), and even the dependence on $\nu$ and $\tE^2$
(equivalently, $\kappa_*=\upi\nu\tE^2$) can be scaled out.  This table can then be used to calculate
$\var(F)$ for any combination of macrolensing, stellar density, and source structure.

Before going into those details, we generalize a change of variables introduced by
\cite{Neindorf2003}.  Associate with each real-valued vector $\vv=(v_x,v_y)$ a complex number
$\tilde v=v_x+iv_y$.  The argument of the inner exponential of eq.~\eqref{eq:charfct} becomes $if_x$
in terms of the meromorphic function
\begin{equation}\label{eq:cforce}
\tilde f(\tilde\xi) \equiv \tE^2\sum_{j=1}^{n}\frac{\tilde\omega_j}{\tilde\xi -\tilde\theta_j}.
\end{equation}
The change of variables $\tilde\xi\to \tilde f$ has jacobian $|d\tilde f/d\tilde\xi|^{-2}$,
so that the integral in eq.~\eqref{eq:charfct}
becomes
\begin{equation}
  \label{eq:cint}
  \nu\iint\left( e^{i f_x} -1 - if_x\right)\left|d\tilde\xi/d\tilde f\right|^2\,df_x df_y\,.
\end{equation}
The derivative $d\tilde\xi/d\tilde f$ is needed as a function of $\tilde f$, which requires
inverting the function \eqref{eq:cforce}.  This leads to a polynomial of degree $n$ in $\tilde\xi$.
The jacobian must be summed over all $n$ roots.  A slight simplification is that the coefficient of
$\tilde\xi^{n-1}$ in the polynomial vanishes because $\sum_j\tilde\omega_j=\sum_j\tilde\theta_j=0$,
so that the $n$ roots for $\tilde\xi$ sum to zero.

In particular, for $n=2$ the roots are 
\[
\tilde\xi = \pm\sqrt{\tilde\theta_1(\tilde\theta_1 +2\tE^2\tilde\omega_1/\tilde f)}. 
\]
Set $\bo=\bo_1=-\bo_2$ and $\dbt=2\bt_2=-2\bt_1$.
Rescale the integration variable to $\tilde
r\equiv u-iv\equiv\tilde f\Delta\tilde\theta/4\tE^2\tilde\omega$.  
Finally, let $\psi$ be the angle between $\bo$ and $\dbt$, and $\Delta\theta\equiv |\dbt|$.
The $2$-point function is then
\begin{subequations}
\begin{equation}\label{eq:Hintro}
\left\langle e^{i\bo\cdot[\bp(0)-\bp(\dbt)]}\right\rangle =
  \exp\left[\frac{\kappa_*(\dt)^2}{4\pi\tE^2}\,H\left(\frac{4\tE^2}{\dt}\bo\right)\right],
\end{equation}
where $H$ as a function of $\bb\Leftrightarrow(\beta\cos\psi,\beta\sin\psi)$ is
\begin{multline}
  \label{eq:Hdef}
H(\bb) \Leftrightarrow H(\beta,\psi)\equiv\\
\iint\frac{\exp[i\beta(u\cos\psi+v\sin\psi)]-1-i\beta(u\cos\psi+v\sin\psi)}
{2(u^2+v^2)^{3/2}\sqrt{(u-1)^2+v^2}}dudv\,.
\end{multline}
\end{subequations}

\section{Numerical Evaluation}
\label{sec:evaluation}

Here we describe the methods used to compute the kernel $\hat J(\bo)$ for the second moment of the
flux [eq.~\eqref{eq:Jhat}].  The first step is to compute and tabulate the function $H(\bb)$
(\S\ref{subsec:inner}).  It is quite easy to incorporate a distribution of stellar masses at
this stage (\S\ref{subsec:mspec}). The final step is to carry out the integration over $\dbt$ in
eq.~\eqref{eq:Jhat}, which is really a double integral since $\dbt$ is a 2-component vector
(\S\ref{subsec:outer}).  As a check on the numerics, we derive asymptotic results for $\omega\tE\ll
1$ and $\omega\tE\gg1$ and interpret these results physically (\S\ref{subsec:physical}).

\subsection{The function $H$}
\label{subsec:inner}

Since $u$ and $v$ in eq.~\eqref{eq:Hdef} are actually scaled versions of the ``force''
\eqref{eq:cforce}, the singularity of the integrand at the origin represents the large area
available to distant stars that exert weak deflections.  Without the subtracted terms in the
numerator, eq.~\eqref{eq:Hdef} would be a two-dimensional Fourier transform.
But without at least the $-1$, the singularity at the origin
would not be integrable.  We considered softening the singularities so that we could separate the
three terms in the numerator and evaluate the first by 2D FFTs.  But instead, like
\cite{Seitz+Schneider1994}, we chose to expand $H(\beta,\psi)$ as a Fourier series in $\psi$:
\begin{align}
  \label{eq:Hseries}
  H(\beta,\psi) =&  \tfrac{1}{2}\upi\int\limits_0^\infty\frac{dr}{r^2} \left[J_0(\beta
    r)-1\right]b_{1/2}^{(0)}(r)\nonumber\\
&+i\upi(\cos\psi)\int\limits_0^\infty\frac{dr}{r^2} \left[J_1(\beta r)-\tfrac{1}{2}\beta
  r\right]b_{1/2}^{(1)}(r)\qquad\nonumber\\
&+\upi\sum_{n=2}^\infty i^n\cos(n\psi)\int\limits_0^\infty
\frac{dr}{r^2} J_n(\beta r) b^{(n)}_{1/2}(r)\nonumber\\
 \equiv& \sum_{n=0}^\infty i^n h_n(\beta)\cos(n\psi)\,.
  \end{align}
This representation has two advantages: (i) $H$ is dominated by the terms for $n\le 2$ at both
large and small $\beta$; (ii) convolution with a stellar mass spectrum, as described below, acts in
the radial direction, i.e. along lines of constant polar angle $\psi$.
The Bessel functions, $J_n$, result from expanding the exponential in \eqref{eq:Hdef}
as a Fourier series in $\psi-\alpha$, where $\alpha=\tan^{-1}(v/u)$.  The
Laplace coefficients (e.g., \citealt{Brouwer+Clemence1961})
\begin{equation}
  \label{eq:lapcoef}
  b^{(n)}_{1/2}(r)\equiv\frac{1}{\upi}\int\limits_0^{2\upi}\frac{\cos n\phi\,d\phi}
{\sqrt{1+r^2-2r\cos\phi}}\ 
= \frac{2}{\upi\sqrt{r}}Q_{n-1/2}\left(\frac{1+r^2}{2r}\right)\,,
\end{equation}
result from expanding the second part of the denominator of eq.~\eqref{eq:Hdef}.  Here $Q_{n-1/2}$
are Legendre functions of the second kind and are calculated in our code by recursion on $n$
starting from expressions for $Q_{-1/2}$ and $Q_{+1/2}$ in terms of complete elliptic integrals, or
by hypergeometric series.  The radial integrals are performed numerically in the complex $r$ plane
with due attention to the branch points at $r=1$.
\begin{figure} 
\includegraphics[width=84mm]{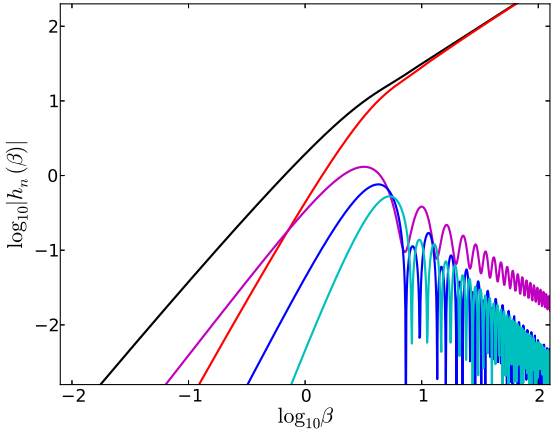}
\caption{The first few azimuthal harmonics $h_n(\beta)$ of the function $H(\bb)$ determining the 2-point
characteristic function [eqs.~\eqref{eq:Hdef} \& \eqref{eq:Hseries}].
From top to bottom at $\beta=1$, these are $n=0,1,2,3,4$.}
\label{fig:fig1}
\end{figure}

To facilitate convolution with a stellar mass function, it is convenient to tabulate the functions
$h_n(\beta)$ on a grid uniformly spaced in $\ln\beta$.  We typically use $\Delta\ln\beta =0.01$ in
the range $10^{-3}\le \beta\le 10^2$.  The first three functions $h_0\ldots h_2$ dominate the
series, though convergence with respect to $n$ is slow at $\beta\sim 1$.  We usually tabulate up to
$n_{\max}=7$, tapering the series to hasten convergence.  These choices are input parameters to our
code.  The functions $h_n(\beta)$ with $n\ge 2$ oscillate in $\beta$ with period $\approx 2\upi$;
this is due to logarithmic singularities of the Laplace coefficients at $r=1$.  In order to resolve
these oscillations on a logarithmic grid, it is necessary that $\beta_{\rm max}\Delta\ln\beta
<\upi$.

For extrapolation in $\beta$ beyond the range of the table,
\begin{align}
  \label{eq:Hlimits}
  H(\beta,\psi) &\to
\tfrac{\upi}{4}\beta^2(\ln\beta +C_0) -\tfrac{\upi}{8}\beta^2\cos2\psi+ O(\beta^3\ln\beta),
\nonumber\\
&\mbox{as }\beta\to 0;
\nonumber\\[1ex]
H(\beta,\psi)&\to -\upi\beta -i\upi(\beta-1)\cos\psi + O(\beta^{-1}), \nonumber\\
&\mbox{as } \beta\to\infty.
\end{align}
The constant $C_0=\gamma-1-\ln8\approx -2.5022$.

\subsection{Incorporating a stellar mass spectrum}
\label{subsec:mspec}

The stellar mass ($M_*$) enters the problem through the square of the Einstein-ring radius, $\tE^2\propto M_*$, in equation
\eqref{eq:charfct} for the $n$-point characteristic function.
 This makes it easy to introduce a spectrum of stellar masses.  Let $f(\log M_*) d\log M_*$ represent the fraction of the
total number of stars that have masses in the logarithmic interval $d\log M_*$.  (Here $\log\equiv\log_{10}$.)
The mean mass is then
\begin{equation*}
  \overline{M}_* = \int\limits_{-\infty}^\infty M_* f(\log M_*) d\log M_*\,.
\end{equation*}
Let $\tE$ be the Einstein-ring radius based on the mean mass,
$\tE^2\equiv 4G\overline{M}_*D_{LS}/c^2 D_LD_S$.  To represent the mass spectrum $f(\log M_*)$ in the 2-point function $H(\bb)$, one
need only make the replacements
by the replacement
\begin{equation}
  \label{eq:hbar}
  h_n(\beta)\to\bar h_n(\beta) = \int\limits_{-\infty}^\infty h_n(10^x\beta)\,f(x)dx\,.
\end{equation}
in the polar expansion \eqref{eq:Hseries}.   The asymptotics \eqref{eq:Hlimits} need to be adjusted accordingly.

In fact, provided $f(x)$ itself is reasonably smooth, integrating over a mass spectrum would accelerate the convergence
of the radial integrals in \eqref{eq:Hseries}, because $\int J_n(10^x\beta r) f(x)dx$ decays exponentially rather than
oscillates when $\log(\beta r)$ is larger than the width of $f(x)$.  This would allow us to integrate entirely on the
real $r$ axis.  However, we choose to tabulate the $h_n$ for a single mass.  The smoothing \eqref{eq:hbar} is performed
very easily after the fact with any desired mass function.  We adopt the log-normal form
\begin{equation}
  \label{eq:fform}
  f(\log M_*) = \frac{1}{\sqrt{2\upi\sigma^2}}\,\exp[-(\log M-\log M_c)^2/2\sigma^2]\,.
\end{equation}
The replacement \eqref{eq:hbar} should leave the mean mass unchanged; this requires that the
characteristic mass $M_c=\overline{M}_*\exp[-(\sigma\ln10)^2/2]$.  In our code, by default,
$\sigma=0.3$ in agreement with the initial mass function recommended by \cite{Chabrier2003} for the
spheroidal component of the Galaxy [$\sigma=0.33\pm0.03$, $M_c=0.22\pm0.05\,M_\odot$], except that
Chabrier replaces the tail of the log-normal function above $0.7\,M_\odot$ with a power law of
roughly Salpeter slope ($x=1.3$).  The assumption is that since the spheroid is an old population,
most of the stars in the tail will have evolved off the main sequence, effectively truncating the
present-day mass function at the turnoff ($0.7\,M_\odot$).  Our code omits this refinement and uses
the log-normal form without truncation.
%With the above parameters, approximately $6.4\%$ of the initial number of stars and $22\%$ of the initial mass lies
%above the turnoff when one uses the uninterrupted log-normal form, versus $12\%$ and $63\%$, respectively, for
%\cite{Chabrier2003}, who assumed Salpeter-like powerlaw with $x=1.3$ instead of $1.35$ above the turnoff.  We consider
%this to be a tolerable error in view of the uncertainty in the present-day mass function of elliptical galaxies (e.g.,
%\citealt{Conroy+vanDokkum2012} and references therein).

\subsection{The outer integral}\label{subsec:outer}

Given the function $H(\bb)$, it remains to compute
\begin{equation}
  \label{eq:Jhat2}
  \hat J(\bo) = \det\bM \int d^2\dbt\, e^{-i\dbt\cdot\bM\cdot\bo}
\exp\left[\frac{\kappa_*\dt^2}{4\pi\tE^2}\,H\left(\frac{4\tE^2\bo}{\dt}\right)\right]\,.
\end{equation}
As written, however, this double integral is not convergent for small $\bo$ and large $\dbt$.
Following \eqref{eq:Hlimits},  $H(\bb)\propto\beta^2\ln\beta$ as
$\beta\to0$,  whence the second exponential above tends to tends to unity as $\omega\tE\to 0$.
The rest of the integrand oscillates with constant modulus, whence $\hat J(\bo)\to
(2\upi)^2\delta^2(\bo)$ as $\bo\to0$.

This is to be expected.  A very extended source will have a Fourier transform $\hat I_0(\bo)$ that
decreases rapidly with increasing $|\bo|$.  Such a source should suffer little microlensing, so that
$\langle F^2\rangle=\langle F\rangle^2$.  But the mean macrolensed flux is $\langle F\rangle =
(\det\bM)^{-1}\hat I_0(\bmath{0})$.  Comparison with eq.~\eqref{eq:Jhat}, shows that the
microlensing kernel must contain exactly the delta function identified in the previous paragraph.

For numerical purposes one must have a convergent expression.  One option is to subtract unity from
the second exponential in eq.~\eqref{eq:Jhat2}.  This removes the problem at small $\omega$ but
creates a similar problem at large $\omega$.  A better tactic is to multiply the integrand by a
broad and smooth window function that gradually tapers to zero at large $\dbt$; this smears the
delta function into a narrow but finite spike centered at $\bo$ without much changing the finite
part of the integral at $\omega>0$.  Instead, we have chosen to evaluate eq.~\eqref{eq:Jhat2} by a
version of Euler summation, which works as follows \citep[e.g.][]{Hardy1949}.  Let $\sum_k (-1)^k a_k$ be an
alternating series in which the terms $\{a_k\}$ have constant sign but may increase,
provided $\lim_{k\to\infty} (a_{k+1}/a_k) = 1$.
Let $S^{(0)}_n= \sum_{k=0}^{n-1} (-1)^k a_k$ be the $n^{\rm th}$ partial sum, and for $m>0$
\begin{equation*}
  S^{(m)}_n =   \frac{1}{2}\left[S^{(m-1)}_n +S^{(m-1)}_{n+1}\right] = 2^{-m}\sum_{j=0}^m {m
  \choose j} S^{(0)}_{n+j}\,.
\end{equation*}
If the sequence $\{S^{(0)}_n\}$ converges, then $\{S^{(m)}_n\}$ converges to the same limit.  But
$\{S^{(m)}_n\}$ may converge when $\{S^{(0)}_n\}$ does not.  For example, if $a_k = k$ then
$\{S^{(m)}_n\}\to -1/4$ for $m\ge 2$, which is the ``correct'' result if this series is
regarded as the limit of $\sum_k k(-x)^k= x(d/dx)(1+x)^{-1}$ as $x\to 1^{-}$.

To apply this, we set $(-1)^k a_k$ equal to the integral~\eqref{eq:Jhat2} restricted to the annulus
\begin{equation*}
  k\frac{\upi}{\mu}\le |\dbt| \le k\frac{\upi}{\mu}\,,\quad\mbox{where }
  \mu\equiv|\bM\cdot\bo|\,.
\end{equation*}
For sufficiently large $k$,
\begin{equation*}
 (-1)^k a_k \sim 2\upi\det\bM \int\limits_{k\upi/\mu}^{(k+1)\upi/\mu}d\dt\, J_0(\mu\dt) (\dt)^{1-\kappa_*\tE^2\omega^2}\,
\end{equation*}
The Bessel function $J_0(z)\sim\sqrt{2/\pi z}\,\cos(z-\upi/4)$ for $z\gg 1$.  Hence $a_k\sim
k^\sigma$ with $\sigma= 1/2\ -\kappa_*\tE^2\omega^2$.  After inspecting the smoothed partial sums
$S^{(m)}_n$ for residual oscillations, our code adds more annuli as needed to enable further
smoothing.  This works reasonably well and automatically discards the delta function at $\bo=0$.
%without which eq.~\eqref{eq:Fsq} yields $\var(F)$ rather than $\langle F^2\rangle$.

\subsection{Limiting behaviors of the kernel and their interpretation}
\label{subsec:physical}
We can check the numerical results against analytical ones for $\omega\tE\ll 1$ and for $\omega\tE\gg1$.
Let $\omega$ be small enough so that it makes sense to replace the second exponential in
\eqref{eq:Jhat2} by the first two terms of its power series.
Discard the leading term (unity), which gives the delta function.
Also ignore the small contribution from the range $0\le\dt\la 4\tE^2\omega$, so that
$H(\bb)$ may be replaced by the top line of
eq.~\eqref{eq:Hlimits}.  After integration over the azimuth of $\dbt$,
\begin{multline}\label{eq:Jsmallom0}
  \hat J(\omega)\approx 2\upi \kappa_* (\omega\tE)^2\det\bM \int_0^\infty d\dt
\\
\times \dt \left[\ln\left(\frac{4\tE^2\omega}{\dt}\right) J_0(\mu\dt)
+\tfrac{1}{2}J_2(\mu\dt)\cos 2\alpha\right]\,,
\end{multline}
where $\alpha$ is the angle between $\bo$ and $\bmath{\mu}\equiv\bM\cdot\bo$.
The standard integral (\citealt{Abramowitz+Stegun1970}, \S11.4.16)
\begin{equation}
  \label{eq:Besselint}
  \int_0^\infty t^\mu J_\nu(t)\,dt = 
2^\mu\,\frac{\Gamma\left(\frac{\nu+\mu+1}{2}\right)}{\Gamma\left(\frac{\nu-\mu+1}{2}\right)}.
\end{equation}
converges only if $\mu+\nu>-1$ and $\mu<\tfrac{1}{2}$ but can be analytically continued via the
right side where that is finite and nonzero.  Logarithms can be inserted by differentiation with
respect to $\mu$.  In particular,
\begin{equation*}
  \int_0^\infty dt\, J_0(t)\, t\ln t \to -1\,,\qquad
  \int_0^\infty dt\, J_2(t)\, t  \to +2.
\end{equation*}
Applying these rules to eq.~\eqref{eq:Jsmallom0} yields
\begin{subequations}\label{eq:Jasymptotics}
\begin{equation}\label{eq:Jsmallom}
  \hat J(\bo)\approx 
4\upi\kappa_*\tE^2\det\bM\ \frac{(\bo\cdot\bM\cdot\bo)^2}{|\bM\cdot\bo|^4}\,,
\quad 0<\omega\tE\ll 1.
\end{equation}
On the other hand, when $\omega\tE\gg 1$, we may approximate $H$ by the
second line of eq.~\eqref{eq:Hlimits}, with the result
\begin{equation}
  \label{eq:Jlargeom}
  \hat J(\bo)\approx 2\upi\det\bM\ 
\frac{\kappa_*\omega}{\left(\kappa_*^2\omega^2+|\bM'\cdot\bo|^2\right)^{3/2}}
\qquad\mbox{if}\quad\omega\tE\gg1.
\end{equation}
\end{subequations}
Here $\bM'\equiv\bM+\kappa_*\bmath{1}$, i.e. the residual of the macrolensing matrix
when the mean stellar convergence is removed.

For a spectrum of stellar masses, eq.~\eqref{eq:Jlargeom} is unchanged, but eq.~\eqref{eq:Jsmallom}
is multiplied by $\overline{M_*^2}/\overline{M}_*^{\,2}$ if $\tE^2$ is based on the mean mass
$\overline{M}_*$.  For a log-normal mass function with dispersion $\sigma$ in $\log M_*$, the correction
factor is $\exp[(\sigma\ln10)^2]$.

\subsubsection{Physical interpretation}
The Fourier transform $\hat I_0(\bo)$ of a compact source has significant power at large values of
its argument, $\bo$.  Therefore, the variance of the microlensed flux from such a source should be
dominated by the asymptotic behavior \eqref{eq:Jlargeom}, which scales with $|\bo|$ as
$\omega^{-2}$.  Suppose that the spatial power spectrum $|\hat I_0(\bo)|^2\sim \mbox{constant}\equiv
C$ for $\omega<\Theta_S^{-1}\equiv\omega_S$, where $\Theta_S$ is the angular size
of the source, and that the power spectrum falls off rapidly for $\omega>\omega_S$.  Then if
$\omega_S\gg\tE^{-1}$, it follows from eq.~\eqref{eq:Jlargeom} that $\var(F)/\langle
F\rangle^2\propto C\log\omega_S$, with a constant of proportionality that depends upon
$\kappa_*$, $\det\bM$, and the angular dependence of $|\bM'\cdot\bo|$.  Hence the flux variance
diverges logarithmically in the limit that the angular size of the source tends to zero, as
expected.

In the opposite limit of an extended source, $\Theta_S\gg \tE$, similar reasoning
based on the small-$\omega$ asymptotics \eqref{eq:Jsmallom} leads to $\var(F)/F_0^2
\propto\omega_S^2\propto\Theta_S^{-2}$.  Why should sources larger than the
Einstein ring of an individual star show any microlensing fluctuations at all?  The
answer is that the stellar convergence, whose average value we denote by
$\kappa_*$, is subject to Poisson fluctuations in the number of stars projected onto the
source:
\begin{equation*}
\frac{\Delta\kappa_*}{\kappa_*}\sim \frac{\Delta N_*}{N_*} \sim N_*^{-1/2}\sim
(\upi\nu\Theta_S^2)^{-1/2}\,.
\end{equation*}
Small fluctuations in the convergence translate linearly to fluctuations in magnification, whose variance
(normalized by the mean magnification) therefore scales as $\Theta_S^{-2}$.

\section{Numerical examples and applications}
\label{sec:examples}

The poster child of microlensed quasars is Q2237+305, variously known as Huchra's Lens after its
discoverer, or the Einstein Cross.  For the brightest of the four visible macroimages,
\cite{Poindexter+Kochanek2010a} estimate that $\kappa$, $\kappa_*$, and $\gamma$ are all close to
$0.4$.  Hence the stars dominate the convergence, and the mean magnification of this image is
$[(1-\kappa-\gamma)(1-\kappa+\gamma)]^{-1}\approx 5$.
\begin{figure}
\includegraphics[width=84mm]{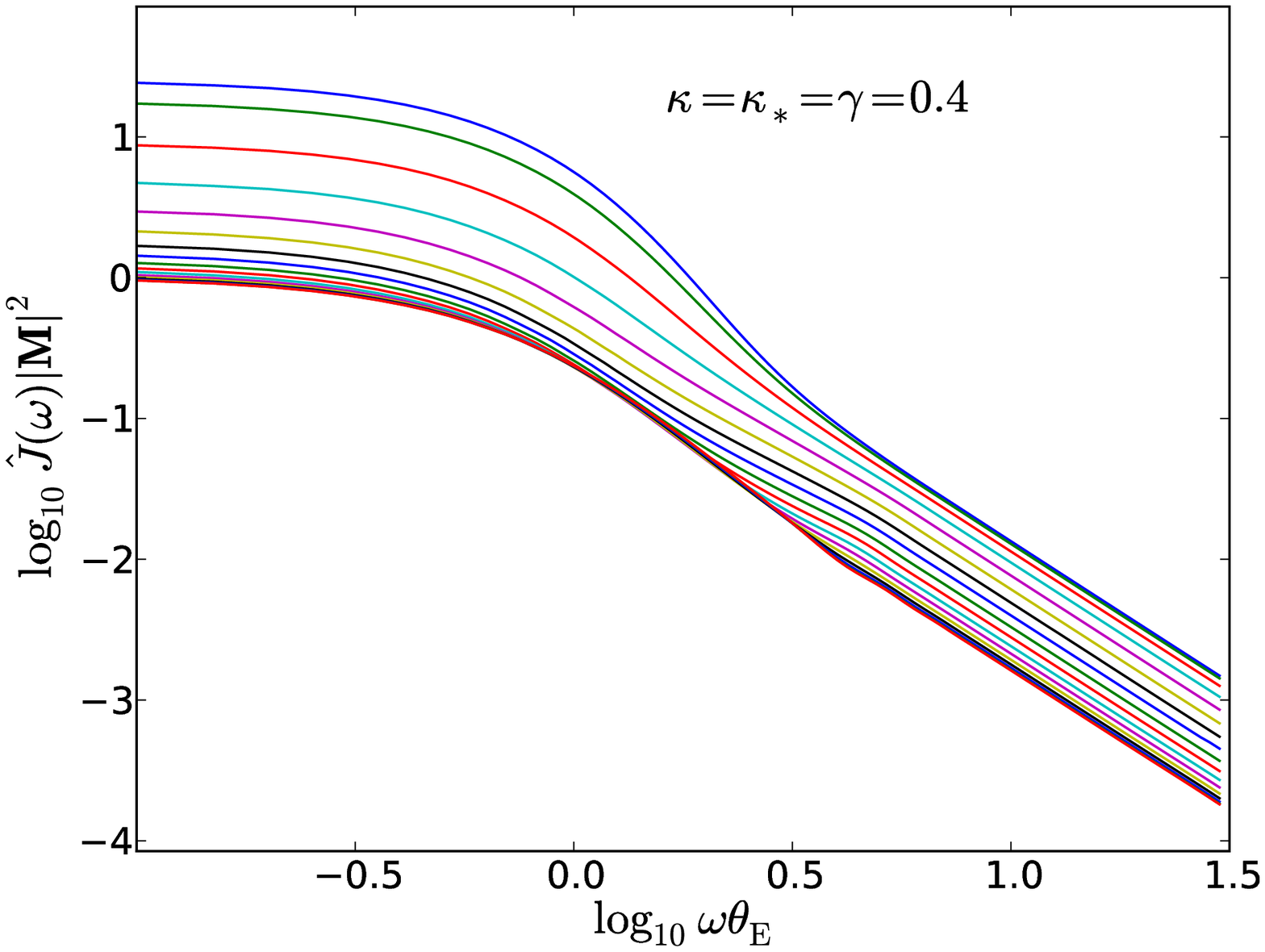}
\caption{The numerically evaluated Fourier-transformed microlensing kernel $\hat J(\bo)$ 
for the macrolensing parameters of Image~A in Q2237+0305 as given by
\citet{Poindexter+Kochanek2010a}.  Constant along each curve is
the angle ($\alpha$) between the spatial wavevector $\bo$ and the major axis
of the macrolensing matrix $\bM$ [eq.~\eqref{eq:Mmatrix}], i.e., the eigenvector
corresponding to the smaller eigenvalue, $1-\kappa-\gamma$.  These values are uniformly
spaced from $\alpha=0$ (\emph{top curve}) to $\alpha=\upi/2$ (\emph{bottom}). }
\label{fig:fig2}
\end{figure}

Figure~\ref{fig:fig2} shows the microlensing kernel numerically computed for these parameters. The
anisotropy of the kernel---its dependence upon the angle $\alpha$ describing the direction of
$\bo$---is due to the anisotropy of the macrolensing matrix, $\bM$.  The symmetries of the kernel
are such that it suffices to calculate $\hat J(\bo)$ for $0\le\alpha\le\upi/2$, where $\alpha$ is
measured with respect to the eigenvector of $\bM$ corresponding to the eigenvalue of smaller
absolute value.
\begin{figure}
\includegraphics[width=84mm]{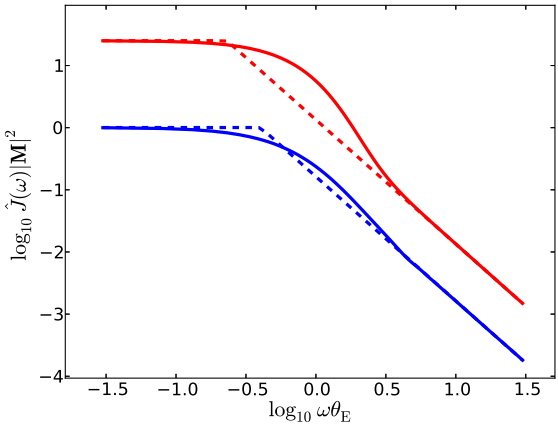}
\caption{Solid lines are the extreme values $\alpha=0$ (\emph{upper curve, in red}) and
  $\alpha=\upi/2$ (\emph{lower, blue}) from Fig.~\ref{fig:fig2}.  Broken dashed lines show
  the asymptotic predictions \eqref{eq:Jasymptotics}.
\label{fig:fig3}}
\end{figure}

%On the vertical scale we plot (the log of) $\hat J(\bo)$ multiplied by $(\det\bM)^2$
%rather than $\hat J(\bo)$ itself.  This is the function that one would use to compute the
%variance of the microlensing magnification, since $\langle F\rangle/F_0 = (\det\bM)^{-1}$:
%\begin{equation}
%  \label{eq:varmag}
%    \var\left(\frac{F}{\langle F\rangle}\right)=
%F_0^{-2}\int \frac{d^2\bo}{(2\pi)^2} |\hat I_0(\bo)|^2
% \left[(\det\bM)^2\hat J(\bo)-(2\pi)^2\delta^2(\bo)\right]\,,
%\end{equation}
%where $F_0$ is the unlensed flux.  The subtracted delta function removes the square of the
%mean microlensing magnification, which by definition is unity because the mean
%magnification is attributed to macrolensing.  Throughout this paper, we use the term
%\emph{microlensing kernel} to refer to $J(\Delta\bT)$, its Fourier transform $\hat
%J(\bo)$, or the combination shown above in square brackets, according to the context.

The logarithmic axes influence the visibility of some details.  There are small-amplitude wiggles in
the curves in the decade $1< \omega\tE<10$ at the level of tens of percents, somewhat more
pronounced for smaller magnifications.  These are vestiges of much stronger oscillations in the
azimuthal harmonics of the $H$ function [Fig.~\ref{fig:fig1}].  The curves in Fig.~\ref{fig:fig2}
for different $\alpha$ are farther apart near $\omega=0$ than as $\omega\to\infty$.  This shows that
$\hat J(\bo)=\hat J(\omega\cos\alpha,\omega\sin\alpha)$ is not 
separable in $\omega$ and $\alpha$.  Although we do not show it here, the inseparability is yet
more striking for the more typical case that $\kappa=\gamma=0.45=10\kappa_*$.

\subsection{Inhomogeneous disks}

\cite[hereafter DA]{Dexter+Agol2011}'s have proposed toy models of inhomogeneous disks, in which the
disk temperature fluctuates around its steady-state value. These models are an interesting test of
our formalism.  They have more structure than simple gaussian source models, yet have spatial
power spectra that are rather easily described.  The fluctuations are spatially
correlated within cells whose width is constant in azimuth and log radius.
Each cell radiates as a black body  at a temperature chosen from a log-normal distribution
($\log\equiv\log_{10}$):
\begin{equation}
  \label{eq:lognormal}
\exp\left\{\left[\log T_r-\log T_{r,0}+ \sigma_T^2\right]^2/\sigma_T^2\right\}
\frac{d\log T_r}{\sqrt\upi\,\sigma_T}.
\end{equation}
Here $T_{0,r}\propto(r/r_s)^{-3/4} $ is the temperature profile of a homogeneous disk with a
constant mass accretion rate, neglecting the inner and outer edges.  The fiducial radius is defined
so that $T_{0,r}=h/k_{\textsc{b}}\lambda$ at $r=r_s$ if $\lambda$ is the wavelength of observation
referred to the disk rest frame.  Following eq.~\eqref{eq:lognormal}, the variance of $\ln T_r$ is
$(\sigma_T\ln10)^2/2$, and the mean of $\ln T_r$ is less than $\ln T_{r,0}$.  The bolometric flux is
the same on average as for the homogeneous disk: $\langle T_r^4\rangle=T_{0,r}^4$.  The narrow-band
luminosity, however, decreases with increasing temperature variance:
$L_\lambda(\sigma_T)=L_\lambda(0)\exp[-\tfrac{8}{9}(\sigma_T\ln10)^2]$.  At the same time, the
apparent size of the source increases: the $n^{\rm th}$ radial moment of the light scales $\propto
L_\lambda r_s^n\exp[\tfrac{4}{9}n(n+1) (\sigma_T\ln 10)^2]$.

The fiducial radius scales with the black-hole mass and accretion rate as $r_s\propto (M\dot
M)^{1/3}$.  However, without reference to $M$ or $\dot M$, one can infer the half-light radius
$r_h(0)$ of a homogeneous disk from the observed narrow-band luminosity, corrected for lensing:
$r_h(0)\propto L_\lambda^{1/2}\lambda^{3/2}$.  The constant of proportionality depends only on
fundamental constants if point on the disk radiates as a black body with temperature scaling as
$r^{-3/4}$.

Based on the temporal variability and optical-to-UV spectra of QSOs, as well as the
microlensing observations, DA conclude that $0.35\la \sigma_T\la0.5$.  Since, as they remark,
$r_h(\sigma_T)/r_h(0)\propto\exp[0.85 (\sigma_T\ln 10)^2]$, it follows that the half-light radii are
larger than those of homogeneous disks by factors of $1.7$ to $3$ when referred to the same
$L_\lambda$.

To apply our methods, we must estimate the spatial power spectra of these models.  We take the disk
to be viewed face on and adopt angular coordinates $\bt\equiv\bmath{r}/D_S$ in the source plane; in
particular, $\theta_s\equiv r_s/D_S$.  Polar coordinates $(\theta,\phi)$ are defined so that
$\bt=(\theta\cos\phi,\theta\sin\phi)$.  We numerically average the Planck function at each radius in
the disk over the temperature distribution \eqref{eq:lognormal}, thus obtaining the mean source $\langle
I(\theta)\rangle_S$.  This is axisymmetric, as is its Fourier/Hankel transform,
\begin{equation}
  \label{eq:Hankel1}
  \langle \hat I(\omega)\rangle_S = 2\upi\int\limits_0^\infty J_0(\omega\theta)\langle
  I(\theta)\rangle_S \,\theta\,d\theta.
\end{equation}
The subscript $S$ serves as a reminder that the average is taken over realizations of the source,
not over the microlensing.  The spatial correlation of the temperature fluctuations---the size of
the cells---plays no role in eq.~\eqref{eq:Hankel1}, because the average $\langle
I(\theta)\rangle_S$ of the surface brightness is computed independently at each point on the disk.

Eq.~\eqref{eq:Fsq} for the microlensing flux variance involves $|\hat I(\omega)|^2$.  Since this is a
random variable in DA's inhomogeneous models, we must average it over realizations of the
source:
\begin{equation}
  \label{eq:sourceavg}
\langle |\hat I(\omega)|^2\rangle_S = |\langle\hat I(\omega)\rangle_S|^2 +  \var_S[\hat I(\omega)]\,.
\end{equation} 
The last term is the Fourier transform of the two-point correlation of the
surface brightness fluctuations,$\langle\Delta I(\bt_1)\Delta I(\bt_2)\rangle$, which vanishes
unless the points $\bt_1$ and $\bt_2$ belong to the same cell.  The correlation depends on
$\phi_1-\phi_2$, $\ln(\theta_1/\theta_2)$, and
$\ln\bar\theta\equiv\tfrac{1}{2}\ln(\theta_1\theta_2)$, but varies more rapidly with the first two
variables than the third if $N_r$, the number of cells per octave in radius, is large.  Thus to an
adequate approximation,
\begin{multline}
\langle\Delta I(\theta_1,\phi_1)\Delta I(\theta_2,\phi_2)\rangle_S\approx \\
\var[I(\theta_1)]\,W_\phi(\phi_1-\phi_2) W_\theta(\ln\theta_1-\ln\theta_2),
\end{multline}
where $W_\phi$ and $W_\theta$ are triangular window functions of width $2\upi/N_\phi$ and
$(\ln2)/N_r$, respectively, $N_\phi$ being the number of cells in azimuth.  Then\footnote{The use of
  $\mbox{sinc\,}x\equiv(\sin x)/x$ is a further approximation but is accurate when $\omega\theta\gg
  \max(|n|,1)$; in the opposite limit, the integrand is negligible anyway.  For numerical
  quadratures we replace $\mbox{sinc}^2(x/2)$ with $\mbox{sech}^2(x/\upi)$, which has the same
  equivalent width and the same effect of suppressing the integrand where $\omega$ is larger than
  the reciprocal of the local cell size, but avoids the sidelobes.}
\begin{multline}
  \label{eq:Pvar}
  \var_S[\hat I(\omega)]\approx \frac{(2\upi)^2\ln 2}{N_\theta N_\phi}
\sum_{n=-N_\phi/2}^{N_\phi/2-1}
\int\limits_0^\infty \var_S[I(\theta)]\, J_n^2(\omega\theta)\\
\times \mbox{sinc}^2\left(\frac{\omega\theta\ln2}{2N_r}\right)\,\theta^3\,d\theta.
\end{multline}
The strength of this term relative to the first term on the right side of \eqref{eq:sourceavg}
increases with $\sigma_T$ at fixed $N_\theta N_\phi$.  DA state that they obtain the best match to
the variability data if the number of cells per octave in radius is in the range $100\la n \la
1500$.  We interpret this $n$ to correspond to $N_\theta N_\phi$ in our notation.  DA do not specify
the aspect ratio of their cells.  We presume that they are roughly square, although because of
differential rotation, one might expect that the correlation length of disk inhomogeneities should
be longer in azimuth than in radius.  We take $N_\theta=10$, $N_\phi=55$, so that
$\Delta\theta=0.11\,{\rm rad}$ and $\Delta\ln r = 0.069$.  We normalize the power spectra to unity
at $\bt=0$: $P(\omega)=\langle|\hat I(\bt)|^2\rangle_S/\langle|\hat I(0)|^2\rangle_S$.  This is not
the same as normalizing by the square of the mean flux, $\langle F\rangle_S=\langle\hat
I(0)\rangle_S$.  When microlensing is measured via flux ratios of macroimages, after correction for
time delays, intrinsic variations of the source flux cancel.
Normalizing the power spectrum to unity at $\bt=0$ is therefore more appropriate than normalizing by
the average flux.  In practice, it does not much matter which normalization one uses for DA's
preferred parameter range.  The ratio $\langle\hat I(0)^2\rangle_S/\langle \hat I(0)\rangle_S^2$
increases with $\sigma_T$ and decreases with $N_\theta N_\phi$.  It is less than $1.04$ for
$\sigma_T\le0.5$ at $N_\theta N_\phi=550$, but rises rapidly for larger temperature dispersions,
reaching $4.53$ at $\sigma_T=0.7$.
\begin{figure}
\includegraphics[width=84mm]{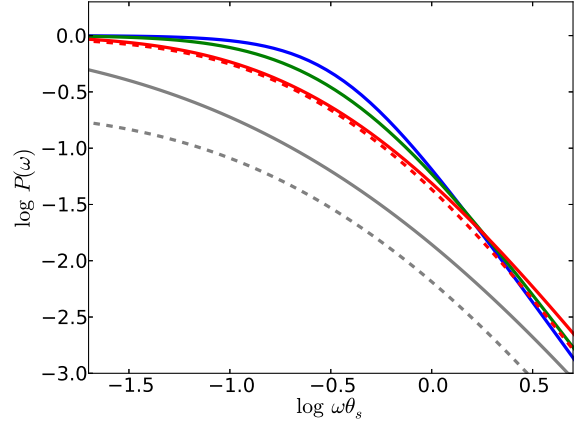}
\caption{Spatial power spectra of inhomogeneous disks with log-normal temperature variations,
  following \citet{Dexter+Agol2011}.  Power spectra are normalized to unity at wavenumber
  $\omega=0$.  Abscissa is scaled by the angular radius $\theta_s$ at which
  $hc/k_{\textsc{b}}T_{\rm eff}=\lambda_{\rm rest}$ in the homogenous ($\sigma_T=0$) disk.  From top
  to bottom at $\log\,\omega\theta_s\approx-0.5$, solid curves correspond to
  $\sigma_T=\{0,0.3,0.5,0.7\}$. Dashed curves show $\langle\hat I(\omega)\rangle_S^2/
  \langle\hat I(0)\rangle_S^2$ rather than $P(\omega)\equiv\langle\hat
  I^2(\omega)\rangle_S/\langle\hat I^2(0)\rangle_S$ for the most variable cases
  ($\sigma_T=0.5,0.7$). }
\label{fig:fig4}
\end{figure}

Figure~\ref{fig:fig4} shows the source power spectra calculated as described for several values of
$\sigma_T$.  As expected, the general effect of increasing
$\sigma_T$ is to suppress the power spectrum at $\omega\theta_s\la 2$, because the mean source
size increases.  However, at least for $\sigma_T\la0.5$, the power spectrum is actually
enhanced at larger $\omega$, probably because the hottest individual cells, which are much smaller
than $\theta_s$, increasingly dominate.  Comparison of the solid and dashed curves shows that for
$\sigma_T\la0.5$, the power spectrum is nearly equal to the square of the Fourier transform of
the mean intensity, which is independent of cell size.  For $\sigma_T\ga0.7$, however, the
cell-to-cell variance is more important, so that the second term in eq.~\eqref{eq:sourceavg} cannot
be neglected.

\begin{figure}
  \includegraphics[width=84mm]{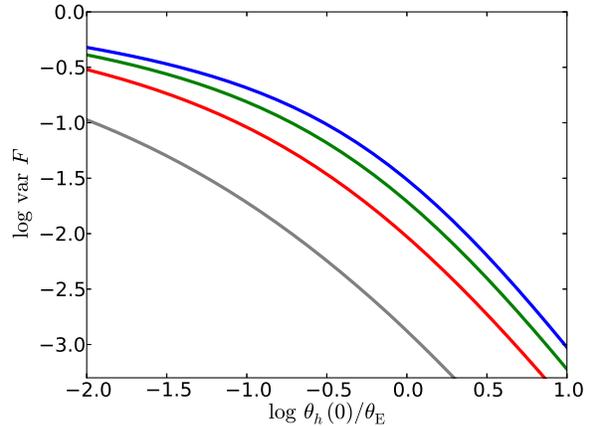}
  \caption{Microlensed flux variance of the inhomogeneous disks from Figure~\ref{fig:fig4} for the
    fiducial microlensing parameters $\kappa=\gamma=0.45$, $\kappa_*=0.045$ versus inferred
    half-light radii $\theta_h(0)$ of homogeneous disks of the same $L_\lambda$.  From top to bottom
    at $\log\,\theta_h(0)/\tE\approx -1$, the curves correspond to $\sigma_T=0,0.3,0.5,0.7$.}
  \label{fig:fig5}
\end{figure}

Figure~\ref{fig:fig5} shows the microlensing variance calculated by applying eq.~\eqref{eq:Fsq} to
the power spectra shown in fig.~\eqref{fig:fig4}.  On the abscissa, the size of the source
is measured not by $\theta_s$, but rather by the half-light radius of a
homogenous disk of equivalent narrow-band luminosity: $\theta_h(0) = 2.44\exp[-(4/9)(\sigma_T\ln
10)^2]\,\theta_s$.  At fixed $\theta_h(0)$, larger $\sigma_T$ makes for a larger true source size
and therefore, as expected, smaller microlensing flux variations.  To gauge the effect of the
temperature fluctuations on the source size inferred from microlensing, it is more appropriate to
consider the horizontal rather than vertical distance between curves, i.e., the dependence of
$\theta_h(0)$ on $\sigma_T$ at fixed $\var F$.  For example, along the curve for $\sigma_T=0$,
variances of $0.05$, $0.1$, $0.2$, and $0.4$ are achieved for $\theta_h(0)/\tE=0.649$, $0.303$,
$0.106$, and $0.0191$; whereas for $\sigma_T=0.5$, these same variances require
$\theta_h(0)/\tE=0.212$, $0.0879$, $0.0266$, and $0.00421$.  The former sizes are larger than the
latter by factors ranging from $3.1$ to $4.5$: these are the factors by which the size of a $\sigma_T=0.5$
source inferred from microlensing variability would exceed the size inferred by fitting a conventional
disk to its narrow-band luminosity.
\begin{figure}
  \includegraphics[width=84mm]{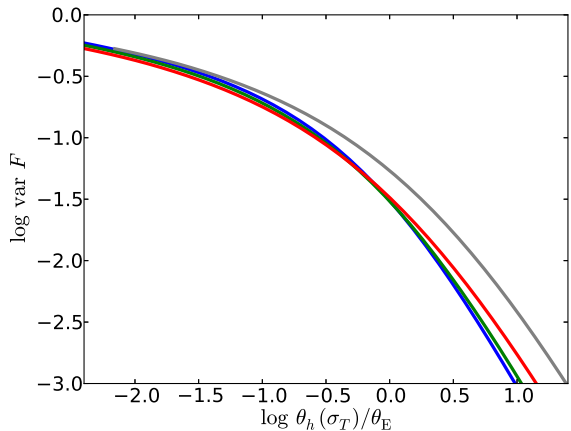}
  \caption{Like Fig.~\ref{fig:fig5}, but plotted against the true half-light radius of the mean
    inhomogenous source.  From top to bottom at $\log\,\theta_h(\sigma_T)/\tE\approx +1$, the curves
    correspond to $\sigma_T=0.7,0.5,0.3,0$.}
  \label{fig:fig6}
\end{figure}
If one plots the microlensing variance against the \emph{true} half-light radius of the
corresponding mean source, then the curves lie almost on top of each other, at least for
$\theta_h\la\tE$ (Fig.~\ref{fig:fig6}).  This confirms the common wisdom that microlensing
variability is relatively insensitive to the details of the source structure at a given half-light
radius.  However, this rule of thumb breaks down for sufficiently wild sources, as the curve for
$\sigma_T=0.7$ shows.

\subsection{Comparison with simulations}

We have tested the predicted flux variance against simulations carried out with an
elementary inverse-ray-shooting code.  The code is our own but its design follows
\cite{Kochanek2004}.  The source and lens plane are taken to be periodic, with periodicity lengths
in the ratio $L_x/L_y=1$ on the source plane and $(1-\kappa+\gamma)/(1-\kappa-\gamma)$ on the lens
plane.  All simulations reported here were performed for $\kappa=\gamma=10\kappa_*=0.45$, so
$(L_x/L_y)_{\rm image plane}\to 10$.  Lensing stars having a log-normal distribution of masses
($\sigma_{\rm mass}=0.3\mbox{ dex}$) are scattered over the lens plane with mean number density
$\kappa_*/\upi\tE^2$.  Their masses are assigned to grid points using a cloud-in-cell method, and
the deflections they cause are computed by Fourier transforms using a particle-mesh (PM)
method.\footnote{\cite{Kochanek2004} used ${\rm P^3M}$ to better resolve short-range forces.
  However, we typically have $\la 10^{-4}$ stars per grid point.  Hence the few
  rays that pass within a cell width of a star carry little light (all rays being
  weighted equally).  They contribute even less to caustics because large deflections entail strong
  demagnification.}  The numbers of cells along each dimension of lens domain are in the ratio
$N_x/N_y\approx\sqrt{L_x/L_y}\to\sqrt{10}$, on the theory that because of the macrolensing shear,
the four microimages\footnote{The softened potentials of our simulated stars create an odd number of
  images, but the central image is demagnified.}  split by an isolated on-axis star have
separations along $x$ and $y$ that lie in this ratio.  The rays form a uniform mesh in the image
plane with four rays per grid cell, corresponding to $4\sqrt{10}$ per (square) cell in the source
plane, for a total of $3.4\times 10^8$ rays in our largest simulations.  It has been argued that a
much larger number of rays per cell is needed for accurate results with the inverse-ray-shooting
method \citep[e.g.][]{Mediavilla+etal2006}.  However, that conclusion is reached on the basis of a
pixel-by-pixel comparison with some analytic solution such as that for an isolated star.  We smooth
the source-plane magnification pattern with gaussians before computing the flux variance, so it is
the number of rays per smoothing length or the number per star that is relevant.

\begin{table*}
  \centering
   \begin{minipage}{140mm}
     \caption{Microlensed flux variances of gaussian sources from simulations and the
       semianalytic method.  Lensing parameters $\kappa=\gamma=0.45$, $\kappa_*=0.045$, \&
       $\sigma_{\rm mass}=0.30\,\mbox{dex}$. See text for column headings.}
\vspace*{10pt}
  \begin{tabular}{rrrrllllll}
    $L_y\ [\tE]$&$N_x$&$N_y$&$N_*$& $\sigma_S=$&0.005$\,\tE$&0.01$\,\tE$&0.02$\,\tE$&
0.04$\,\tE$&0.08$\,\tE$
\\ \hline\hline\\
&&&&\multispan{2}{\hfill\it Resolution tests\hfill}&&&&\\[1ex]
51.2&  8192&2592&477&$\dotfill$&  0.56(1)& 0.48(1)& 0.40(1)& 0.32(1)& 0.236(6) \\
51.2&16384&5184&477&$\dotfill$&  0.63(2)& 0.55(2)& 0.46(1)& 0.37(1)& 0.28(1)\\[2ex]
&&&&\multispan{2}{\hfill\it Domain-size tests\hfill}&&&&\\[1ex]
 25.6&   4096&1296&119&$\dotfill$&  0.58(3)& 0.50(2)& 0.42(2)& 0.33(2)& 0.25(2)\\
 51.2&   8192&2592&477&$\dotfill$&  0.56(1)& 0.48(1)& 0.40(1)& 0.32(1)& 0.236(6)\\
102.4&16384&5184&1907&$\dotfill$&0.602(7)&0.525(7)&0.440(6)&0.353(6)&0.267(5)\\
\\\hline\\
\multispan{5}{{\bf Semi-analytic}\dotfill}& 0.6134 & 0.5286 & 0.4440 & 0.3598 & 0.2767
  \end{tabular}
  \end{minipage}
  \label{tab:sims}
\end{table*}

Table~\ref{tab:sims} compares the variances obtained from our simulations to the semi-analytic
predictions for gaussian sources\footnote{I.e., the unlensed surface brightness is
  $I_0(\bt)\propto\exp(-\theta^2/2\sigma_S^2)$.}  with dispersions ranging from $\sigma_S=0.005\tE$
to $0.08\tE$.  The first column lists the periodicity length in the $y$ direction, which is common to
the source and lens/image planes; $L_x=L_y$ in the source plane and $=10L_y$ in the image plane.
The second and third columns give the number of grid points in each dimension, and the fourth, the
number of stars.  In columns 5-9, the numbers in parentheses represent the uncertainty in
the last digit of the estimated variance based on the spread in the results of four simulations sharing
the same parameters listed in the first four columns but differing in the random assignments of
positions and masses to the stars.

The flux variances estimated from the simulations are clearly noisy, but the agreement with theory
tends to improve both with resolution and with domain size.  The improvement with domain size is in
part merely a reduction in noise due to the larger numbers of stars, but there appears also to be a
systematic trend in the mean values, which suggests that long-range forces are important.  In a
periodic domain, the forces exerted by each star must also be periodic and hence cannot follow the
correct scaling $\bp\propto|\Delta\bt|^{-1}$ at separations $\Delta\bt$ larger than about half the
periodicity length.  We have found that the variances are sensitive to the approximation chosen for
the force kernel.  The results shown here were obtained by constructing the force kernel in
coordinate space and windowing the exact kernels for the $x$ and $y$ components of the deflection
with $\cos(\upi\Delta\theta_x/L_x)$ and $\cos(\upi\Delta\theta_y/L_y)$, respectively.  More
sophisticated simulation algorithms, such as the inverse-polygon method of
\cite{Mediavilla+etal2006}, would likely improve the rate of convergence with resolution.  It would
be interesting to see whether they also help the convergence with domain size.

\section{Summary}

Motivated in part by discrepancies between the angular sizes of QSOs inferred from microlensing and
those expected from disk theory, we have developed a practical method for computing the variance of
the microlensed fluxes of angularly extended sources of arbitrary structure.  The method requires as
inputs the spatial power spectrum of the unlensed source, the shear and convergence provided by the
smooth mass distribution of the lensing galaxy, and the mean number density and mass function of the
lensing stars.  We have written and made available for download a small suite of codes that accepts
these inputs and calculates the microlensing variance.  The mathematical formulation of the method
is described above in sufficient detail to allow an interested user to understand the workings of
our codes.  Further improvements in efficiency and accuracy are doubtless possible.  The
semi-analytic method has been tested against a simple inverse-ray-shooting simulation code and
compared with the published results of \cite{Dexter+Agol2011} for their toy models of highly
inhomogeneous disks.

Perhaps the most important lesson learned from these tests is that significant numerical effort---in
terms of the numbers of pixels and simulated stars, as well as the number of independent trials---is
necessary to obtain good agreement between the flux variance estimated from the simulations and that
calculated semi-analytically by the method developed here.  Doubtless a cleverer simulation method
could get by with fewer rays, but the need to simulate a large region of the lens planes, $\ga
10^2\tE$ on a side, seems inescapable, at least when the macrolensing magnification is large ($\sim
10$), as it typically is in present lensing surveys.

\section*{Acknowledgments}
We thank Jason Dexter for help with understanding \cite{Dexter+Agol2011}.

%\newpage
\bibliographystyle{mn2e}
\bibliography{lensing}

\end{document}